# Gain and plasmon dynamics in negative-index metamaterials

By Sebastian Wuestner, Andreas Pusch, Kosmas L. Tsakmakidis,
Joachim M. Hamm and Ortwin Hess*

*Department of Physics, South Kensington Campus, Imperial College London,
London SW7 2AZ, UK*

Photonic metamaterials allow for a range of exciting applications unattainable with ordinary dielectrics. However, the metallic nature of their meta-atoms may result in increased optical losses. Gain-enhanced metamaterials are a potential solution to this problem, but the conception of realistic, three-dimensional designs is a challenging task. Starting from fundamental electrodynamic and quantum-mechanical equations we establish and deploy a rigorous theoretical model for the spatial and temporal interaction of lightwaves with free and bound electrons inside and around metallic (nano-) structures and gain media. The derived numerical framework allows us to self-consistently study the dynamics and impact of the coherent plasmon-gain interaction, nonlinear saturation, field enhancement, radiative damping and spatial dispersion. Using numerical pump-probe experiments on a double-fishnet metamaterial structure with dye molecule inclusions we investigate the build-up of the inversion profile and the formation of the plasmonic modes in the low-$Q$ cavity. We find that full loss compensation occurs in a regime where the real part of the effective refractive index of the metamaterial becomes more negative compared to the passive case. Our results provide a deep insight into how internal processes affect the over-all optical properties of active photonic metamaterials fostering new approaches to the design of practical loss-compensated plasmonic nanostructures.

## 1. Introduction

An important recent advance in electromagnetics and optics has been the conception and development of composite media featuring negative electromagnetic parameters [1–3]. Negative effective permittivities, permeabilities and refractive indices can allow for cancelling the effect that the corresponding positive constants of standard dielectrics have on the propagation or localization of lightwaves. This often leads to a variety of important optical parameters being either dramatically reduced or enhanced, entering unusual regimes beyond the reach of ordinary materials. For instance, in the field of plasmonics [4] the negative (real part of the) permittivity $\varepsilon_m$ of a metal can be used to cancel the contribution made by the positive permittivity $\varepsilon_d$ of a dielectric to the wavevector $k$ of a surface plasmon polariton $[k_{\text{SPP}}^2 \sim \varepsilon_m \varepsilon_d/(\varepsilon_m + \varepsilon_d)]$ at a metal-dielectric interface. At the frequency where $\varepsilon_m \approx -\varepsilon_d$ the wavevector $k$ (wavelength $\lambda$) dramatically increases (decreases), thereby allowing for sub-wavelength guiding and manipulation of light signals. Likewise, a negative-refractive-index lens can allow for the cancellation of the phase accumulated by a lightwave in a positive-index medium and for the restoration of the amplitude of evanescent waves, giving rise to high-resolution imaging [5, 6]. A similar rationale, the use of a 'negative' parameter to cancel a corresponding 'positive' one, is exploited in a whole range of extraordinary applications enabled by such media, including negative-index based 'invisibility' cloaking [7] and stopping of light in metamaterial and plasmonic waveguides [8–10].

---

* Author for correspondence (o.hess@imperial.ac.uk).

It has been shown that metamaterials can be engineered for almost any regime of the electromagnetic spectrum, from d.c. [11] to ultraviolet frequencies [12]. This is usually achieved either by using doped (metal-like) semiconductors [13, 14] or by arranging sub-wavelength metallic elements inside or on top of a host dielectric medium [1–3]. The role of these inclusions is to act as the meta-atoms of the engineered medium. Circulating currents and oscillating electric charges on the meta-atoms in response to an incident external lightwave create an effective magnetization and polarization of the metamaterial, thereby altering its observed permittivity $\varepsilon$, permeability $\mu$ and refractive index $n$ – to the point of even allowing for an engineered medium with $\text{Re}(n) < 0$.

The metallic nature of the meta-atoms unavoidably leads to high dissipative losses at optical wavelengths. In the negative-index regime, in particular, a metamaterial must necessarily be dispersive in order to ensure that the field energy density inside it remains positive [1, 2]. Owing to these reasons, the absorption coefficients for passive, optical negative-index metamaterials are usually high and can exceed the bulk gain coefficients of common active media by more than an order of magnitude. The level of absorption losses that currently characterizes plasmonic metamaterials limits the practical deployment in the important applications outlined above, and for this reason there have recently been intensified efforts to identify solutions to this problem [15–28].

In a series of interesting recent works [16–20] it has been shown theoretically that field enhancement due to plasmonic resonances can substantially (by more than a factor of ten) increase the effective gain coefficient associated with the active metamaterial compared to that of bulk active media. This effective gain enhancement makes it possible to fully compensate the dissipative losses in optical metamaterials by using available gain media (dyes or semiconductor quantum wells/dots). Indeed, a recent experimental work has observed amplification of lightwaves passing through an active double-fishnet metamaterial [21], while significant and ongoing experimental progress is also currently under way pertaining to gain-enhanced magnetic metamaterials [25–27].

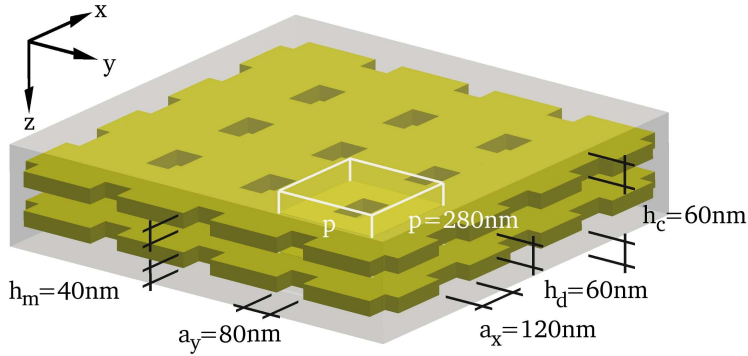

Figure 1. Illustration of the double-fishnet structure with its associated geometric parameters. A square unit cell with side-length $p$ is highlighted in white. The two perforated silver films are embedded in a dielectric host material (translucent) infiltrated with dye molecules. (Online version in colour.)

The negative effective refractive index of a standard (passive or active) double-fishnet metamaterial [29–31] is attainable in the direction perpendicular to the plane defined by the holes of one of the fishnets, and can be retrieved using reflection/transmission measurements [18–21, 32]. Parallel to this plane the metamaterial is spatially dispersive and therein its effective refractive index is not defined [33]. The cross-sections of the holes also play a crucial role in determining the coupling of the photons existing inside the metamaterial to the continuum of the radiation modes outside it, i.e. in determining its radiation losses and thereby affecting its $Q$-factor. This non-trivial interplay of (in-plane) spatial dispersion, radiative damping and reflectivities at the two planar ends of a fishnet metamaterial, together with the interaction of plasmonic excitations with the gain medium over a three-dimensional topology that includes sharp edges and corners, calls for the development of a general and rigorous theoretical/numerical framework that can reliably grasp and model the interaction of light with these structures.

It is the objective of this work to detail, evaluate and deploy a self-consistent model to carry out numerical studies on the time-dependent influence of available gain media on the optical properties of negative-index metamaterials. The geometry of the double-fishnet metamaterial [29, 30] under consideration is given in figure 1. While our gain model may approximate a large variety of different gain materials, such as quantum wells and quantum dots, we here assume molecules of the dye Rhodamine 800 embedded in a dielectric host material [19, 21] with refractive index $n_h = 1.62$. Note that in practice the Rhodamine 800 molecules photo-bleach after a few minutes, thereby preventing long-term operation [21]. Despite this limitation, dye molecules can be used for proof-of-principle studies and for comprehending and probing the plasmon/gain dynamics in active plasmonic metamaterials. Resorting to such a configuration is also motivated by the fact that the required integration of e.g. crystalline semiconductor gain media (e.g., quantum-wells) between non-crystalline or amorphous metals in order to maximize the gain-enhancement effect might be a difficult experimental task.

Here, we directly model the electromagnetic fields and their resonant interaction with the free-electron plasma of the metals and the transitions of the bound electrons in the gain medium [19, 34–38]. The dynamics of the photonic fields, the resonant excitation of (localized) surface plasmons (SPs) and the temporal evolution of the carrier inversion in the gain inclusions emerge naturally from a set of fundamental dynamical equations. These are Maxwell's curl equations in their full-vectorial form supplemented by adequate local response models to capture the coherent response of the metallic structures patterned on the nanoscale and the absorption/emission characteristics of the chromophores embedded in the dielectric spacer. The model self-consistently describes the processes that influence the spatial and temporal evolution of excited plasmonic modes inside the metamaterial: excitation by the injected pump- and probe-fields, interaction with the gain medium by stimulated emission and absorption processes, and temporal decay due to ohmic and radiative damping. Making sure that experimental conditions are reflected as accurately as possible, we make predictions about the optical properties of the considered nanostructures.

The article is organized as follows. Section 2 introduces the theory and models on which we base our investigations. Specifically, we therein detail aspects of the modelling of the surface plasmon opto-dynamics, the two-level optical Bloch equations and the four-level gain model used for the description of the laser dye. The methods used in our numerical analyses are presented in §3: We give details on the finite-difference time-domain (FDTD) method, the setup for the transmission/absorption measurements and the method by which we extract the effective electromagnetic parameters of the active double fishnet. In addition, we verify the accuracy of our implementation of the four-level gain model. In §4 we present the results of the modelling analysis focusing, in particular, on the spatio-temporal dynamics of the inversion during the pump process, the evolution of the plasmonic mode at the probe wavelength, and the retrieved effective refractive index and magnetic permeability of the active double fishnet. Finally, §5 summarizes the paper, drawing the main conclusions from the present study.

## 2. Theory

In this section we present the fundamental equations for and models employed in our study of dynamical processes inside metamaterials. By taking into account the relevant length-scales we provide a formulation of Maxwell's equations for the evolution of the classical electromagnetic fields inside optical metamaterials (§2*a*). We implement a dynamical model for surface plasmons, which are formed by the interaction of electromagnetic waves with the free electron plasma in metals (here silver) (§2*b*), and for the interaction of Rhodamine 800 molecules with coherent electromagnetic fields (§2*d*) based on the optical Bloch equations (§2*c*).

(*a*) *Electromagnetic fields in media: Maxwell's equations and response models*

Maxwell's equations govern the temporal evolution of the classical electromagnetic fields $\mathbf{E}(\mathbf{r},t)$ and $\mathbf{B}(\mathbf{r},t)$ in the presence of charges and currents. By averaging over (small) volumes $\Delta V(\mathbf{r})$ one finds the representation of Maxwell's equations in media, which in SI units reads

$$\frac{\partial \mathbf{D}}{\partial t} = \nabla \times \mathbf{H} - \mathbf{J}\,, \quad \nabla \cdot \mathbf{D} = \rho\,, \\ \frac{\partial \mathbf{B}}{\partial t} = -\nabla \times \mathbf{E}\,, \qquad \nabla \cdot \mathbf{B} = 0\,. \tag{2.1}$$

Here, as a consequence of the spatial averaging, $\rho(\mathbf{r},t)$ accounts only for the net charge density while $\mathbf{J}(\mathbf{r},t)$ is the injected current density. $\mathbf{D}(\mathbf{r},t)$, the displacement current, and $\mathbf{H}(\mathbf{r},t)$, the magnetic field, contain contributions from currents induced within the media.

For metamaterials the wavelength $\lambda$ has, by definition, to be considerably larger than the diameter $L$ of the structural elements (meta-atoms). A spatially resolved description of these elements on the other hand requires $L$ to be far larger than the diameter of the volume of averaging, $\Delta l$. In the optical regime this implies that we have to read equations (2.1) as mesoscopic field equations where $\lambda \gg L \gg \Delta l$. In our calculations $\Delta l$ takes a typical value of $\sim 5$ nm.

In the absence of free charges ($\rho = 0$) and injected currents ($\mathbf{J} = 0$) it is generally sufficient to solve the two curl equations in (2.1), which preserve the divergence of the fields at all times. However, due to the introduction of new fields, equations (2.1) are not a closed system any more and need to be supplemented by constitutive equations $\mathbf{H}[\mathbf{B}]$ and $\mathbf{D}[\mathbf{E}]$, describing how charges and currents locally respond to the electromagnetic fields contributing to the media polarization and magnetization. On the mesoscopic scale we assume non-magnetic media, i.e. $\mathbf{B} = \mu_0 \mathbf{H}$, and write the constitutive equation for the displacement current as

$$\mathbf{D} = \varepsilon_0 \varepsilon \mathbf{E} + \mathbf{P} \tag{2.2}$$

where $\mathbf{P} = \mathbf{P}[\mathbf{E}]$ is introduced as a dynamic polarization response and $\varepsilon$ as a static non-resonant background dielectric permittivity. By inserting this definition into the curl equations of (2.1) one obtains two coupled equations

$$\varepsilon_0 \varepsilon \frac{\partial \mathbf{E}}{\partial t} = \nabla \times \mathbf{H} - \frac{\partial \mathbf{P}}{\partial t}\,, \tag{2.3a}$$

$$\mu_0 \frac{\partial \mathbf{H}}{\partial t} = -\nabla \times \mathbf{E}\,, \tag{2.3b}$$

governing the spatial and temporal evolution of electromagnetic waves in interaction with media.

Generally, $\mathbf{P}[\mathbf{E}]$ has a non-instantaneous and nonlinear functional dependence on the exciting electric field. In frequency-domain, the linear case $\mathbf{P}(\omega) = \varepsilon_0 \chi(\omega) \mathbf{E}(\omega)$ is associated with a number of fundamental electronic response models. For example, poles $\chi(\omega) = d/(c - ib\omega - a\omega^2)$ with coupling strength $d/a$, pole frequency $\sqrt{c/a}$ and relaxation constant $b/a$ can, depending on the choice of parameters, either represent the Drude model for free electrons ($a = 1$, $c = 0$), the Debye model for a relaxing dipole ($a = 0$, $c = 1$) or the Lorentz model for a resonant excitation of bound electrons ($a = 1$). Linear frequency-domain response models transform into linear ordinary differential equations for the time-dependent polarization locally driven by the electric field strength. In particular, the discussed pole equation translates into

$$a\frac{\partial^2 \mathbf{P}}{\partial t^2} + b\frac{\partial \mathbf{P}}{\partial t} + c\mathbf{P} = \varepsilon_0 d \mathbf{E}. \tag{2.4}$$

Beyond the modelling of elementary responses, superpositions of functional responses $\mathbf{P}[\mathbf{E}] = \sum_i \mathbf{P}_i[\mathbf{E}]$ can be used to approximate experimental data over a broad range of frequencies.

### (b) Time domain modelling of surface plasmons

The dynamical interaction of electromagnetic fields with the free electron plasma of a metal is the fundamental mechanism for the excitation and propagation of surface plasmons (SPs) on metal-dielectric interfaces. The simplest response model that, when coupled to Maxwell's equations, produces these excitations is the Drude model. In the context of this work we are interested in the realistic modelling of SPs excited and localized on the thin perforated silver films embedded within the metamaterial structure. The model we adopt overlays a Drude response (D) with two Lorentzian resonances (L1,L2) to locally (at each point) create a response, which approximates experimentally measured thin-film data within the wavelength range $\lambda = 300$–$800$ nm. In this approach, the electric field $\mathbf{E}(\mathbf{r}, t)$ dynamically drives three polarizations $\mathbf{P}_i(\mathbf{r}, t)$, each represented by a separate differential equation of the type (2.4),

$$\frac{\partial^2 \mathbf{P}_\mathrm{D}}{\partial t^2} + \gamma_\mathrm{D} \frac{\partial \mathbf{P}_\mathrm{D}}{\partial t} = \varepsilon_0 \omega_\mathrm{D}^2 \mathbf{E}, \tag{2.5a}$$

$$\frac{\partial^2 \mathbf{P}_\mathrm{L1}}{\partial t^2} + 2\gamma_\mathrm{L1} \frac{\partial \mathbf{P}_\mathrm{L1}}{\partial t} + \omega_\mathrm{L1}^2 \mathbf{P}_\mathrm{L1} = \varepsilon_0 \Delta\varepsilon_\mathrm{L1} \omega_\mathrm{L1}^2 \mathbf{E}, \tag{2.5b}$$

$$\frac{\partial^2 \mathbf{P}_\mathrm{L2}}{\partial t^2} + 2\gamma_\mathrm{L2} \frac{\partial \mathbf{P}_\mathrm{L2}}{\partial t} + \omega_\mathrm{L2}^2 \mathbf{P}_\mathrm{L2} = \varepsilon_0 \Delta\varepsilon_\mathrm{L2} \omega_\mathrm{L2}^2 \mathbf{E}, \tag{2.5c}$$

with parameters as listed in table 1. The time derivative of the sum of the polarizations $\mathbf{P}[\mathbf{E}] = \mathbf{P}_\mathrm{D}[\mathbf{E}] + \mathbf{P}_\mathrm{L1}[\mathbf{E}] + \mathbf{P}_\mathrm{L2}[\mathbf{E}]$ with appropriate frequency response is then coupled back into the wave-equation (2.3).

In a spatially resolved calculation the auxiliary differential equations (2.5) have to be computed at each point where silver is located. It has been shown that the numerical implementation of this model (based on a finite-difference time-domain scheme) yields excellent agreement between numerical simulations and experimental measurements of resonant SP excitation on single silver nanocubes [39].

Table 1. Parameters for the thin-film silver model taken from the supplementary information of [39]. The response, using a superposition of one Drude (D) and two Lorentzian resonances (L1, L2), is represented in time-domain by equations (2.5).

|  | non-resonant permittivity | plasma frequency $\omega_\mathrm{D}$ (rad s$^{-1}$) | relaxation rate $\gamma_\mathrm{D}$ (s$^{-1}$) |
| --- | --- | --- | --- |
| Drude (D) | 1.17152 | $1.39604 \times 10^{16}$ | 12.6126 |

|  | resonance strength $\Delta\varepsilon_\mathrm{L}$ | resonance frequency $\omega_\mathrm{L}$ (rad s$^{-1}$) | resonance half-width $\gamma_\mathrm{L}$ (s$^{-1}$) |
| --- | --- | --- | --- |
| Lorentzian (L1) | 2.23994 | $8.25718 \times 10^{15}$ | $1.95614 \times 10^{14}$ |
| Lorentzian (L2) | 0.222651 | $3.05707 \times 10^{15}$ | $8.52675 \times 10^{14}$ |

Inside metamaterials, SP modes are excited when the propagating electromagnetic wave dynamically interacts with free electrons of the metallic nanostructure. Non-propagating (localized) SP resonances, which can be excited inside a plasmonic resonator, are inherently lossy (ohmic losses) and couple to free-space photonic modes with a characteristic radiation damping rate. We emphasize that the resonant excitation, evolution and temporal decay of both, propagating and non-propagating (localized) SP modes emerge naturally from the presented model.

### (c) *Optical Bloch equations for a two-level system*

It is central to the description of gain dynamics in active media to capture the quantum mechanical nature of the coherent resonant interaction between electromagnetic fields and bound electrons in atoms or molecules, namely the processes of stimulated emission and absorption. The basic quantum mechanical model for these processes is that of an electronic two-level system coupled to the electric field in electric dipole approximation.

By applying Liouville's equation to the density operator $\hat{\rho}$ of the two-level system one obtains the optical Bloch equations, which are formally equivalent [40] to the Bloch equations for nuclear magnetization [41]. The full quantum mechanical formulation dictates that the density operator $\hat{\rho}$ dynamically interacts with the quantized electric field modes. In the semiclassical approximation one assumes a classical electric field $\mathbf{E}(\mathbf{r}, t)$, thereby neglecting spontaneous emission but retaining the coherent processes of the light-dipole interaction. The dynamics of the matrix elements of $\hat{\rho}$, the complex dynamical polarization $\rho_{12}$ and the real-valued occupation densities $\rho_{11}$ and $\rho_{22} = 1 - \rho_{11}$, are then given by

$$\frac{\partial \rho_{12}}{\partial t} = \frac{\partial \rho_{21}^*}{\partial t} = -\left(i\omega_r + \Gamma\right)\rho_{12} - i\frac{\boldsymbol{\mu} \cdot \mathbf{E}}{\hbar}\left(\rho_{22} - \rho_{11}\right), \tag{2.6a}$$

$$\frac{\partial \rho_{22}}{\partial t} = -\frac{\partial \rho_{11}}{\partial t} = i\frac{\boldsymbol{\mu} \cdot \mathbf{E}}{\hbar}\left(\rho_{12}^* - \rho_{12}\right) = \frac{2\boldsymbol{\mu} \cdot \mathbf{E}}{\hbar}\operatorname{Im}\left(\rho_{12}\right), \tag{2.6b}$$

where $\omega_r = (E_2 - E_1)/\hbar$ is the transition resonance frequency between the two energy levels $E_2$ and $E_1$ and $\boldsymbol{\mu} = \langle 1|\mathbf{r}|2\rangle$ is the dipole matrix element of the electronic transition. Additionally, a phenomenological term with dephasing rate $\Gamma$ has been introduced to account for decoherence of the polarization $\rho_{12}$.

When working with classical real-valued fields it is usually more practical to transform the first order complex equations for $\rho_{12}$ into second order real-valued differential equation for $\operatorname{Re}(\rho_{12})$ [37, 38]. This can be achieved by applying some simple algebraic manipulations, first by separating equation (2.6a) into real and imaginary parts:

$$\frac{\partial \operatorname{Re}(\rho_{12})}{\partial t} = -\Gamma \operatorname{Re}(\rho_{12}) + \omega_r \operatorname{Im}(\rho_{12}), \tag{2.7a}$$

$$\frac{\partial \operatorname{Im}(\rho_{12})}{\partial t} = -\Gamma \operatorname{Im}(\rho_{12}) - \omega_r \operatorname{Re}(\rho_{12}) - \frac{\boldsymbol{\mu} \cdot \mathbf{E}}{\hbar}\left(\rho_{22} - \rho_{11}\right). \tag{2.7b}$$

By taking the time derivative of equation (2.7a) and substituting equation (2.7b) we get

$$\frac{\partial^2 \operatorname{Re}(\rho_{12})}{\partial t^2} = -\Gamma \frac{\partial \operatorname{Re}(\rho_{12})}{\partial t} + \omega_r \left[-\Gamma \operatorname{Im}(\rho_{12}) - \omega_r \operatorname{Re}(\rho_{12}) - \frac{\boldsymbol{\mu} \cdot \mathbf{E}}{\hbar}\left(\rho_{22} - \rho_{11}\right)\right]. \tag{2.8}$$

Equation (2.7a) is now used to eliminate $\operatorname{Im}(\rho_{12})$ and all terms containing $\operatorname{Re}(\rho_{12})$ are collected on the lhs. This results in the desired second order differential equation for $\operatorname{Re}(\rho_{12})$:

$$\frac{\partial^2 \operatorname{Re}(\rho_{12})}{\partial t^2} + 2\Gamma \frac{\partial \operatorname{Re}(\rho_{12})}{\partial t} + \left(\omega_r^2 + \Gamma^2\right)\operatorname{Re}(\rho_{12}) = -\omega_r \frac{\boldsymbol{\mu} \cdot \mathbf{E}}{\hbar}\left(\rho_{22} - \rho_{11}\right). \tag{2.9}$$

The corresponding equation for the temporal evolution of the occupation densities is obtained from equation (2.6b) by eliminating $\operatorname{Im}(\rho_{12})$ using (2.7a),

$$\frac{\partial \rho_{22}}{\partial t} = -\frac{\partial \rho_{11}}{\partial t} = \frac{2\boldsymbol{\mu} \cdot \mathbf{E}}{\hbar \omega_r}\left[\frac{\partial \operatorname{Re}(\rho_{12})}{\partial t} + \Gamma \operatorname{Re}(\rho_{12})\right]. \tag{2.10}$$

The optical Bloch equations include saturation nonlinearity via the coupling of the occupation probabilities $\rho_{11}$ and $\rho_{22}$ to the differential equation of the coherence $\operatorname{Re}(\rho_{12})$. They can be employed as a simple model for a saturable absorber but they can not self-consistently describe amplification by an optically pumped gain medium as they lack a pump-channel to maintain inversion. Nonetheless, the derivation of the semiclassical equations (2.9) and (2.10) from quantum mechanical principles delivers the vital ingredients for the time-domain modelling of coherent optical transitions in multi-level systems.

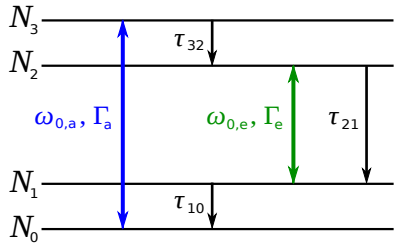

Figure 2. Sketch of the four-level system and its parameters. (Online version in colour.)

(*d*) *Four-level system as gain model for optically pumped laser dyes*

An optically pumped gain system, such as fluorescent dye, coherently absorbs energy at a pump wavelength and emits into the signal field at a longer wavelength. A self-consistent model for these processes therefore requires at least two optical transitions, for the pump $(0 \leftrightarrow 3)$ and for the signal $(1 \leftrightarrow 2)$ in an arrangement as depicted in figure 2. These transitions, each associated with a polarization equation similar to (2.9), can be phenomenologically coupled to a four-level system by adding non-radiative carrier relaxation processes $3 \to 2$ and $1 \to 0$. Optical transitions between the subsystems $(0,3)$ and $(1,2)$ are considered to be dipole forbidden.

The following equations of motion describe the temporal evolution of the carrier densities $N_j$ of the four-level system shown in figure 2:

$$\frac{\partial N_3}{\partial t} = \frac{1}{\hbar \omega_{r,a}} \left( \frac{\partial \mathbf{P}_a}{\partial t} + \Gamma_a \mathbf{P}_a \right) \cdot \mathbf{E}_{\text{loc}} - \frac{N_3}{\tau_{32}}, \qquad (2.11a)$$

$$\frac{\partial N_2}{\partial t} = \frac{N_3}{\tau_{32}} + \frac{1}{\hbar \omega_{r,e}} \left( \frac{\partial \mathbf{P}_e}{\partial t} + \Gamma_e \mathbf{P}_e \right) \cdot \mathbf{E}_{\text{loc}} - \frac{N_2}{\tau_{21}}, \qquad (2.11b)$$

$$\frac{\partial N_1}{\partial t} = \frac{N_2}{\tau_{21}} - \frac{1}{\hbar \omega_{r,e}} \left( \frac{\partial \mathbf{P}_e}{\partial t} + \Gamma_e \mathbf{P}_e \right) \cdot \mathbf{E}_{\text{loc}} - \frac{N_1}{\tau_{10}}, \qquad (2.11c)$$

$$\frac{\partial N_0}{\partial t} = \frac{N_1}{\tau_{10}} - \frac{1}{\hbar \omega_{r,a}} \left( \frac{\partial \mathbf{P}_a}{\partial t} + \Gamma_a \mathbf{P}_a \right) \cdot \mathbf{E}_{\text{loc}}. \qquad (2.11d)$$

The polarization densities $\mathbf{P}_a = \mathbf{P}_a(\mathbf{r},t)$ of the transition $0 \leftrightarrow 3$ and $\mathbf{P}_e = \mathbf{P}_e(\mathbf{r},t)$ of the transition $1 \leftrightarrow 2$ are driven by the local electric field $\mathbf{E}_{\text{loc}}(\mathbf{r},t)$ according to

$$\frac{\partial^2 \mathbf{P}_i}{\partial t^2} + 2\Gamma_i \frac{\partial \mathbf{P}_i}{\partial t} + \omega_{0,i}^2 \mathbf{P}_i = -\sigma_i \Delta N_i \mathbf{E}_{\text{loc}}, \quad i = \text{a,e}. \qquad (2.12)$$

Here, $\Delta N_a(\mathbf{r},t) = N_3(\mathbf{r},t) - N_0(\mathbf{r},t)$ is the inversion of the pump transition and $\Delta N_e(\mathbf{r},t) = N_2(\mathbf{r},t) - N_1(\mathbf{r},t)$ the inversion of the probe transition. We also introduce the resonance frequencies $\omega_{0,i} = (\omega_{r,i}^2 + \Gamma_i^2)^{1/2}$ and a phenomenological coupling constant $\sigma_i$. We further note that in equations (2.11) and (2.12) we use the local electric field at the location of the dipoles which, in a dielectric, can differ substantially from the electric field that appears in the mesoscopically averaged Maxwell's equations. In Lorentz approximation the local and averaged fields inside a dielectric are connected via $\mathbf{E}_{\text{loc}} = [(2 + \varepsilon')/3]\,\mathbf{E}$ [42]. Details of the numerical implementation of the complete model based within the FDTD framework (see §2*a*) are provided in Appendix A.

The coherent effects of stimulated emission and absorption, as well as the nonlinear effects of gain saturation and depletion of both transitions are inherently included in the model equations. For a sustained pumping these equations will provide local gain experienced by a signal propagating through the structure, which depending on the strength of feedback can lead to either laser or amplifier operation. If the (round-trip) gain in the system is high enough to overcome internal and radiation losses a small signal field probing the system will be enough to kick-start the lasing process [22]. The onset of lasing can therefore be triggered by external stimulation even though spontaneous emission in itself is not included in the semiclassical model.

In our work we investigate how laser dye molecules embedded in the dielectric host material affect the properties of a plasmonic metamaterial structure. Fundamentally, the inhomogeneously broadened absorption and emission lines are formed by a multitude of transitions associated with different vibrational states. The presented four-level model with its homogeneously broadened transitions therefore can only provide a good approximation in situations where effects of spectral hole burning can be neglected. This is the case for the numerical experiments carried out here, as we consistently use spectrally broad, short pulses which do not selectively act on individual vibrational states. It is important to note that the dynamical variables $N_j$ and $\mathbf{P}_i$ are thought to emerge from a mesoscopic summation over the occupation densities and polarizations of the individual molecules, i.e. the density of dye molecules is given by $N = \sum N_j$. Assuming randomly oriented dipole moments the sum over molecular polarizations will result in a polarization driven in the direction of the electric field vector, producing the isotropic coupling of equation (2.12).

The parameters of the model, the coupling constants $\sigma_i$, the resonance halfwidths $\Gamma_i$ and the relaxation times $\tau_{ij}$, are empirically chosen to approximate the experimentally measured characteristics of the dye under consideration. The Lorentzian lineshapes of our model, in particular, can only provide a rough approximation to the measured lineshapes of the absorption and emission lines, which arise from a combination of homogeneous broadening due to decoherence and inhomogeneous broadening due to the multitude of vibrational states replaced by two effective transitions. For Rhodamine 800 considered in this work we have chosen the values shown in table 2 based on experimental data extracted from [43].

## 3. Methods

For solving equations (2.3) numerically we extend the finite-difference time-domain (FDTD) method in an approach where the discretized Maxwell's equations are integrated together with the equations for the media polarization (§3*a*). In §3*b* we give details on the setup of the numerical transmission/reflection experiment, followed by a description of the optical parameter retrieval method (§3*c*). The accuracy and stability of the four-level dynamical gain model are validated by performing numerical tests (§3*d*).

Table 2. Parameters of the four-level system.

|  | wavelength $\lambda_{0,i}$ (nm) | coupling constant $\sigma_i$ (C$^2$/kg) | resonance half-width $\Gamma_i$ (fs$^{-1}$) |
|---|---|---|---|
| emission line (e) | 710 | $1.05 \times 10^{-8}$ | 1/20 |
| absorption line (a) | 680 | $1.35 \times 10^{-8}$ | 1/20 |

|  | relaxation time $\tau_{21}$ (ps) | relaxation time $\tau_{32} = \tau_{10}$ (fs) | dye molecule density $N$(cm$^{-3}$) |
|---|---|---|---|
|  | 500 | 100 | $6 \times 10^{18}$ |

(*a*) *Solving the wave equation: the finite-difference time-domain method*

The finite-difference time-domain (FDTD) method [44] is based on the discretization of equations (2.3) in time and space with field components defined on a grid according to

$$F|_{i-i_0,j-j_0,k-k_0}^{n-n_0} = F((i-i_0)\Delta x, (j-j_0)\Delta y, (k-k_0)\Delta z, (n-n_0)\Delta t). \quad (3.1)$$

The offsets $(i_0, j_0, k_0, n_0)$ of the spatio-temporal grid with indices $(i, j, k, n)$ are chosen such that the field components of the discretized electric and magnetic fields are shifted by half-steps against each other in an arrangement known as Yee's staggered grid [45]. The advantage of this is that all first-order derivatives in Maxwell's curl equations (2.3) become second-order accurate with respect to the chosen spatial and temporal step sizes. The electric field components are centred at the time-step $n$ while the magnetic field components are centred at $n + 1/2$. This implies that a complete field update can be performed in two half-steps using a leapfrog integration scheme. Numerical stability is subject to the Courant-Friedrichs-Lewy (CFL) condition which dictates that the largest possible single time-step is limited to the spatial step times $1/\sqrt{3}$ for the three-dimensional case.

The FDTD method enables us to not only describe the spatio-temporal evolution of the fields in arbitrary dielectric geometries but also to include the dynamic electronic responses in form of induced polarization currents – even in the nonlinear regime. For this purpose we need to locally solve sets of (auxiliary) differential equations to obtain the dynamic (nonlinear) polarization. This is achieved by integrating the respective discretized equations numerically alongside the electromagnetic fields taking into account the appropriate spatial and time-centring of the field components. At each point in space we then couple back the induced polarization current into Maxwell's equations (2.3).

For the presented study of an optically pumped active metamaterial we employ sets of auxiliary differential equations to model how the pump and the probe fields dynamically interact with the free electron plasma of the metal (§3*b*) and the bound electrons of the gain inclusions via stimulated emission and absorption (§3*d*).

### (*b*) Numerical method to calculate transmission/reflection spectra

The numerical experiments are based on transmission/reflection calculations in which we inject short pulses with planar wavefronts for both the pump- and probe-fields from the front of the computational domain. The incident pulses dynamically interact with the spatially resolved features of the double-fishnet metamaterial, resulting in parts of the electromagnetic field being reflected, absorbed and transmitted. We record time-series of the reflected and transmitted fields at each point of the front- and back-plane of the simulation domain, perform Fourier-transforms and determine the complex transmission $t(\omega)$ and reflection $r(\omega)$ coefficients. From these we also calculate the spectral energy fluxes expressed in the transmission $T(\omega) = |t(\omega)|^2$, reflection $R(\omega) = |r(\omega)|^2$ and absorption $A(\omega) = 1 - T(\omega) - R(\omega)$.

In our simulations we employ a combination of perfectly-matched layer boundary conditions (PMLs) at the front from where we inject the pulses and the back where the transmitted energy leaves the box-shaped three-dimensional computational domain. In both lateral directions the unit cell is enclosed between periodic boundary conditions.

Like most other implementations (e.g. Berenger's PMLs) the uniaxial PMLs used here are susceptible to late-time instabilities. This effect is generally enhanced when combining PMLs with periodic boundary conditions. Therefore, special care has been taken during the geometry setup in choosing a sufficiently large distance between the metamaterial structure and the PMLs to prevent these instabilities from interfering with the transmission/reflection measurements.

### (*c*) Optical parameter retrieval method

For the extraction of the effective electromagnetic parameters of the active double fishnet we adopt the methodology of [32], which draws upon making an analogy between a metamaterial slab and a dielectric Fabry-Pérot. Using this method, for an arbitrary metamaterial structure, the experimentally or numerically obtained transmission and reflection coefficients can be used to either calculate the permittivity and permeability or the refractive index and impedance that a dielectric slab with these coefficients would have. Retrieving these param-eters allows one to examine the dispersion and signs of the real and imaginary parts of the homogenized refractive index, permittivity and permeability of the metamaterial. The procedure is as follows: First, the complex transmission and reflection coefficients are recovered from plane-wave measurements at normal incidence. Disregarding the internal structure, the measured coefficients enter a calculation to retrieve the effective parameters of the homogenized metamaterial slab of thickness $d$, which may faithfully reproduce the scattering spectra of the actual metamaterial. For example, the formula for the so-extracted effective refractive index is given by

$$n_{\text{eff}} = \pm \frac{\arccos\left[\left(1 - r^2 + t'^2\right)/(2t')\right] + 2\pi m}{kd} \ . \quad (3.2)$$

with the reflection coefficient $r$ and the modified transmission coefficient $t' = \exp(ikd)t$.

We note that the solution for the effective refractive index at a wavelength $\lambda = 2\pi/k$ is multi-branched as it is based on the inversion of the formula for calculation the transmission/reflection coefficients from the optical parameters of a slab. This means, that the sign ($\pm$) and the integer value $m$ in (3.2) have to be uniquely determined. The value $m$ stems from the phase uncertainty of $2\pi$ in the complex scattering parameters. For a passive structure the ($\pm$) sign can immediately be determined by demanding $\text{Im}(n_{\text{eff}}) \geq 0$, and for micrometer-size thin slabs the phase integer $m$ is usually close to zero.

For thicker slabs, the phase integer is often determined by demanding that the real part of the refractive index is a continuous function of the wavelength. When gain is present, in which case $\text{Im}(n_{\text{eff}})$ might also be negative, a useful way of determining the sign and the $m$-factor in equation (3.2) is by identifying the $n_{\text{eff}}$ values for which causality and Kramers-Kronig relations are obeyed [46]. This last complementary methodology has been applied successfully in [19–21].

The above method generally leads to accurate and physically legitimate results [47] if the conditions of homogeneity and non-bianisotropy are met, i.e. if the dimensions of the meta-atoms are considerably smaller than the wavelength of the probing light and the structure is symmetric from both sides. Here, the homogenization of the fishnet metamaterial is performed for normal incidence with the wave-vector along the $z$-direction (see figure 1), perpendicularly to the $xy$-plane of the holes [33]. For non-normal incidence the effective electromagnetic parameters become nonlocal and spatially dispersive. In addition, the dimensions of the cross sections of the holes on the $xy$-plane affect the coupling of the excited (bright) plasmonic modes to the continuum of the radiation modes outside the metamaterial, and its $Q$-factor. Quasi-static homogenization models that neglect the presence of significant spatial dispersion at oblique angles may, thus, lead to unsound conclusions in assessing whether the gain-enhanced double fishnet is absorbing, transparent, amplifying or lasing.

(*d*) *Validation of the implementation of the four-level gain model*

In order to validate the numerical implementation of the four-level system, we compare results from numerical simulations to analytic expressions. We perform these simulations on a thin slab (in comparison to the wavelength) of host material in which the gain molecules are embedded. First, the dispersion of the refractive index of the active material system is calculated in the linear regime. In a second test we look at the steady-state occupation of the upper emission state $\langle N_2 \rangle$ under continuous pumping at the absorption frequency.

(i) *Small-signal gain and refractive-index dispersion of the four-level system*

The refractive index of a combined host and active medium can be extracted from numerical experiments using the retrieval method outlined in §3*c*. In order to validate our numerical framework we first investigate a situation in which the retrieved parameters can be compared with analytic results. The numerical measurement is performed at very low electric field amplitudes ensuring that the occupation densities stay constant during the short duration of the probe pulse, i.e. $\Delta N_i =$ const. Accordingly, the dipole response is linear and analytic values can be derived from equation (2.12) by taking into account the linear connection between the refractive index and the susceptibility given by $n'(\omega) \approx [n_\mathrm{h}^2 + \chi_\mathrm{a}'(\omega) + \chi_\mathrm{e}'(\omega)]^{1/2}$ and $n''(\omega) \approx [\chi_\mathrm{a}''(\omega) + \chi_\mathrm{e}''(\omega)]/(2n_\mathrm{h})$. In writing these two approximate equations, we assume that the host refractive index $n_\mathrm{h}$ is much larger in magnitude than the complex-valued susceptibilities $\chi_i(\omega)$. The frequency dependent dipole response $\mathbf{P}_i(\omega)$, which is given by

$$\mathbf{P}_i(\omega) = -\frac{\left(\omega_{0,i}^2 - \omega^2 + \mathrm{i}2\Gamma_i\omega\right)}{\left(\omega_{0,i}^2 - \omega^2\right)^2 + 4\Gamma_i^2\omega^2}\sigma_i\Delta N_i \mathbf{E}_\mathrm{loc}(\omega) , \quad (3.3)$$

determines the susceptibility of the medium according to $\mathbf{P}_i(\omega) = \varepsilon_0 \chi_i(\omega)\mathbf{E}(\omega)$ and it follows that

$$n'(\omega) \approx \left[n_\mathrm{h}^2 - \frac{2 + n_\mathrm{h}^2}{3\varepsilon_0}\sum_{i=\mathrm{a,e}}\frac{\omega_{0,i}^2 - \omega^2}{\left(\omega_{0,i}^2 - \omega^2\right)^2 + 4\Gamma_i^2\omega^2}\sigma_i\Delta N_i\right]^{1/2} , \quad (3.4a)$$

$$n''(\omega) \approx -\frac{2 + n_\mathrm{h}^2}{6n_\mathrm{h}\varepsilon_0}\sum_{i=\mathrm{a,e}}\frac{2\Gamma_i\omega}{\left(\omega_{0,i}^2 - \omega^2\right)^2 + 4\Gamma_i^2\omega^2}\sigma_i\Delta N_i . \quad (3.4b)$$

The factor $(2 + n_\mathrm{h}^2)/3$ comes from the local electric field with $\mathbf{E}_\mathrm{loc} = [(2 + \varepsilon')/3]\,\mathbf{E}$ (see §2*d*).

Figure 3 compares the refractive index extracted from numerical simulations to the two equations above for the fixed occupation densities $\Delta N_\mathrm{a} = -N_0 =$ $-N/2$ and $\Delta N_\mathrm{e} = N_2 = N/2$. The agreement between simulation and theory is very good, with errors of less than 0.1% in the real part of the refractive index. These errors can be attributed to numerical inaccuracies in the phase gathered during propagation of the pulse and are highly dependent on the grid resolution (here 5 nm). The imaginary part in figure 3*b* shows a positive absorption peak at $\lambda_\mathrm{a} = 680$ nm and a negative emission peak at $\lambda_\mathrm{e} = 710$ nm. This test validates the small-signal response and confirms the ability of the retrieval method to give correct results for amplifying media.

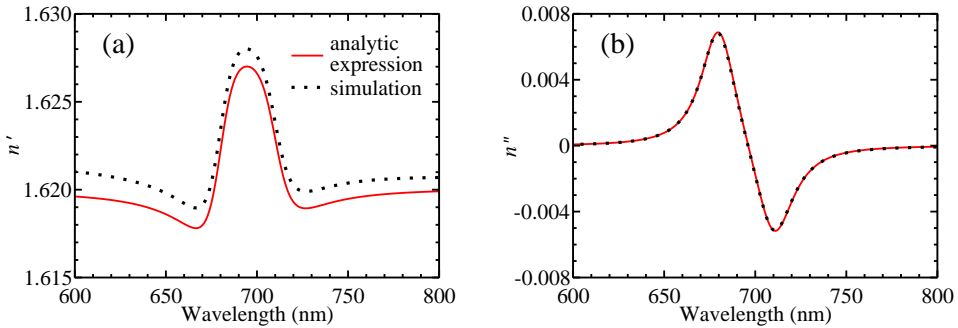

Figure 3. Comparison between numerical simulations (black dotted line) and the analytic equations (3.4) (red solid line) for the real (a) and imaginary (b) part of the refractive index of a thin slab of gain material with $\Delta N_\mathrm{a} = -N_0 = -N/2$ and $\Delta N_\mathrm{e} = N_2 = N/2$. The parameters of the four-level gain material are given in table 2 and the host refractive index is $n_\mathrm{h} = 1.62$. (Online version in colour.)

(ii) *Steady-state occupation under continuous plane wave excitation*

This second test probes the steady-state occupation of the four-level system under continuous plane-wave excitation at the absorption frequency $\omega_{0,\mathrm{a}}$. In the steady-state limit the average occupation densities in equations (2.11) are constant, i.e. $\langle \partial N_j / \partial t \rangle = 0$, and only dependent on the intrinsic parameters of the gain system.

Assuming the overlap of the emission line with the absorption frequency to be negligible, the steady-state of equations (2.11) is given by the set of equations

$$\langle N_0 \rangle = N - \langle N_3 \rangle - \langle N_2 \rangle - \langle N_1 \rangle, \tag{3.5a}$$

$$\langle N_3 \rangle = \langle N_1 \rangle = \frac{\tau}{\tau_{21}} \langle N_2 \rangle, \tag{3.5b}$$

$$\langle N_2 \rangle = \frac{\tau_{21}}{\hbar \omega_{r,\mathrm{a}}} \left\langle \left( \frac{\partial \mathbf{P}_\mathrm{a}}{\partial t} + \Gamma_\mathrm{a} \mathbf{P}_\mathrm{a} \right) \cdot \mathbf{E}_\mathrm{loc} \right\rangle, \tag{3.5c}$$

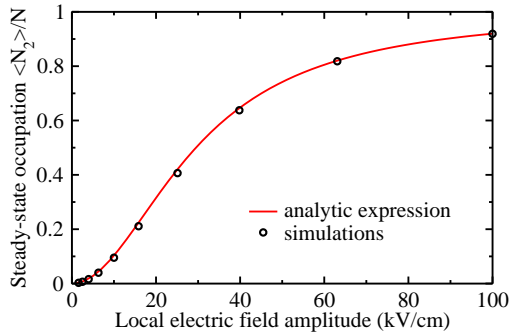

Figure 4. Comparison between simulation results (black circles) and the analytic expression (3.9) for the normalized steady-state occupation density of the upper emission state $\langle N_2 \rangle/N$ of the four-level system (red solid line) when pumping at the absorption resonance $\lambda_{0,a} = 680$ nm. The parameters of the four-level gain material are given in table 2 and correspond to a saturation field amplitude of $E_{\text{sat}} \approx 29.4$ kV/cm. (Online version in colour.)

assuming $\tau = \tau_{32} = \tau_{10}$. The steady-state of equation (2.12) is

$$\left\langle \left(\frac{\partial \mathbf{P}_a}{\partial t} + \Gamma_a \mathbf{P}_a\right) \cdot \mathbf{E}_{\text{loc}} \right\rangle = -\frac{\sigma_a}{2\Gamma_a}(\langle N_3 \rangle - \langle N_0 \rangle)\langle \mathbf{E}_{\text{loc}}^2 \rangle \tag{3.6}$$

$$= -\frac{\sigma_a}{2\Gamma_a}\left[\left(1 + \frac{3\tau}{\tau_{21}}\right)\langle N_2 \rangle - N\right] \frac{E_{\text{loc}}^2}{2}. \tag{3.7}$$

with the amplitude of the local field $E_{\text{loc}}$. Equations (3.5a) and (3.5b) were used to obtain equation (3.7).

Defining the saturation field amplitude $E_{\text{sat}}$ as

$$E_{\text{sat}}^2 = \frac{4\hbar\omega_{r,a}\Gamma_a}{\sigma_a \tau_{21}}, \tag{3.8}$$

equations (3.5c) and (3.7) can be simplified and the occupation of the upper emission state (3.5c) can be rewritten as

$$\langle N_2 \rangle = \frac{N}{1 + 3\tau/\tau_{21} + |E_{\text{sat}}/E_{\text{loc}}|^2}. \tag{3.9}$$

The steady-state occupation of $\langle N_2 \rangle$ is measured at an arbitrary position in a thin slab of gain material. Because the thickness of the slab is chosen to be much shorter than the pump wavelength and absorption length of the dye, the electric field and also the occupation densities can be assumed to be uniform inside the medium. Figure 4 shows the good agreement between numerical steady-state results of the four-level system and the values predicted by equation (3.9). In this specific case where the relaxation time $\tau_{21}$ is much larger than $\tau$, the steady-state inversion $\langle \Delta N_e \rangle$ is approximately equal to $\langle N_2 \rangle/N$ and accordingly a maximum value close to 1 can be achieved at high field amplitudes. At the saturation field amplitude $E_{\text{sat}}$ the normalized occupation density is $\langle N_2\rangle/N \approx 0.5$, which is confirmed by figure 4.

The excellent agreement between numerical results and the analytical expressions of the two presented tests highlight the validity and good accuracy in the numerical implementation of the four-level system.

## 4. Results

In this section we deploy the numerical model detailed in the previous sections to study the plasmon and gain dynamics in a double-fishnet metamaterial with embedded laser dye (see figure 1), and we examine how the macroscopic optical properties of this structure emerge. We conduct numerical pump-probe experiments by first pumping the embedded laser dye with a 2 ps intense pulse polarized along the $x$-direction (see figure 1). After a 7 ps delay we inject a weak pulse of 12 fs duration, again polarized along the $x$-direction, that probes the optical properties of the metamaterial in the linear regime.

In §4*a* we investigate how the gain inversion evolves during the pump process and how it is affected by the field profile of the plasmonic mode at the pump wavelength. Section 4*b* explains the process by which loss compensation is achieved at the probe wavelength and presents the transmission and absorption characterizing both the passive and the active configuration. Finally, §4*c* shows the retrieved effective refractive indices that confirm the attainment of loss compensation in the negative-index regime.

### (*a*) *Pump process: Dynamic build-up of gain inversion*

We first investigate the pump process that leads to the inversion of the gain system necessary for compensation of the metal losses at the probe wavelength. Figure 5*a* reports an exemplary result for the nonlinear dynamics of the gain medium's occupation densities when loss compensation is achieved. During the short duration of pumping, a strong nonlinear response can be observed in the occupation densities. Starting with all molecules in the ground state 0 a large fraction is excited by absorption of the pump light. From the upper absorption state 3 they relax to the upper emission state 2 via non-radiative decay on a time-scale of $\tau = 100$ fs, leaving only a transient population in state 3. The strong pump pulse, though spectrally sharp, also weakly couples to the wide emission line and leads to the temporary population of the lower emission state 1. After approximately 4 ps, when the pump pulse exits the structure, the dynamics slow down considerably and are then governed by non-radiative decay. Accordingly, the very slow decay of $N_2$ is caused by the relaxation $\tau_{21} = 500$ ps. We find that by this process the pump pulse produces a long-lived population inversion $\Delta N_e = N_2 - N_1$ on the emission line of the dye molecules, which can be harvested by a probe pulse.

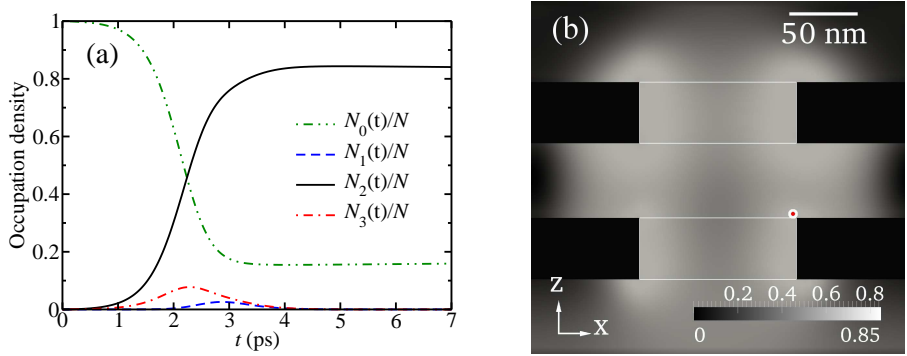

Figure 5. (a) Time dependence of the occupation densities of the gain medium recorded at a position of high inversion given by a red dot in (b). (b) Snapshot of the occupation inversion after the pump process in a plane inside the double-fishnet unit cell. The white rectangular outlines highlight the positions of the holes and the rectangular black areas are the silver films. (Online version in colour.)

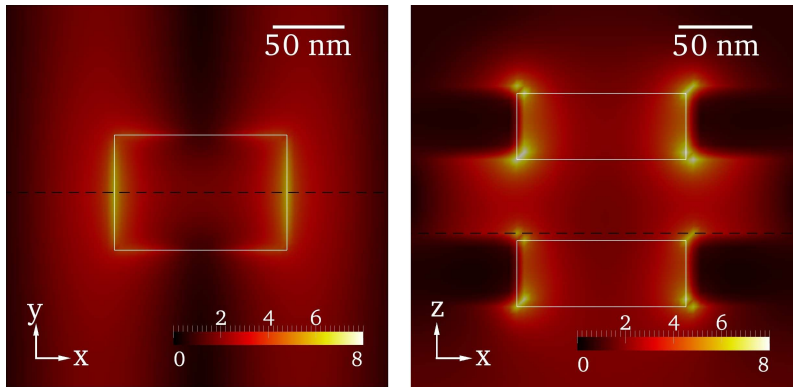

Figure 6. Electric field enhancement at the pump wavelength in two perpendicular planes in a unit cell of the double fishnet. Higher field enhancements are in lighter colours. The white rectangles highlight the position of the holes and the black dashed lines show the intersection of the two planes. (Online version in colour.)

The results shown in figure 5$a$ have been recorded at a position of high inversion highlighted by the red dot in figure 5$b$. In the latter figure we show the spatial distribution of the occupation inversion after the pump pulse has left the system. It can be seen that the occupation inversion is highly spatially non-uniform, with the highest values of 0.85 located close to the metal surfaces. The distribution of the inversion resembles the spatial profile of the excited mode at the pump wavelength presented in figure 6. In particular, a higher local pump-field intensity leads to a higher inversion of the gain material limited by saturation that reduces the imprint of the field enhancement on the occupation inversion.

The strong spatial variation of the pump field originates from the resonant excitation of plasmonic modes, concentrating in particular at the metal-dielectric interfaces and, even stronger, at the sharp corners of the holes. We see from figure 6 that in the regions of high electric field the enhancement relative to the free-space electric field amplitude reaches values close to a factor of 10.

The evolution of the population inversion $\Delta N_e$ can be followed from the successive snapshots depicted in figure 7. While the spatial profile of the pump-field does not vary in time, the profile of the population inversion undergoes significant changes. Initially, it localizes tightly at the edges and corners of the holes but at later stages it progressively expands in both planes shown in figure 7.

Due to the field enhancement the molecular dipoles experience a different pump-strength at different locations. In some of these locations (e.g. around the corners of the holes) the local pump drives the local inversion earlier into saturation than in other locations (e.g. around the centre). This is the reason why the inversion shown at later times in figure 7 does not keep on increasing with time at the points of saturation but continues to broaden in space.

### (*b*) *Probe process: Plasmonic resonator with gain*

Having examined the pump process we now turn our attention to the interplay between the gain inversion and the plasmonically enhanced local electric fields, which results in loss compensation of the probe pulse at the emission wavelength of the dye. Looking at the field enhancement at this wavelength for the passive metamaterial structure, shown in the upper row of figure 8, one may observe the regions where the electric field of the plasmonic mode is highest. These regions are around the edges and the corners of the holes, predominantly between the metal films. In order to achieve maximum efficiency in amplifying the probe pulse these should also be the regions of the highest local gain coefficients, i.e. the regions with the highest local gain inversion. Indeed, we see from the last row of figure 7 that the matching between inversion and field enhancement arising from the plasmonic resonances is well achieved due to the fact that both probe and pump, having the same polarization, couple to the same spectrally broad plasmonic resonance. This enables efficient amplification of the whole plasmonic mode (independent of position) at the emission wavelength because the energy fed locally into the whole mode by the gain medium is proportional to the local gain coefficient multiplied by the local intensity of the electric field [48]. We find from the lower row in figure 8 that with this process the electric field is enhanced by a factor of up to 1.7 compared to the passive case (upper row). Note also that the gain profile does not significantly impact on the shape of the plasmonic mode at the emission wavelength. This confirms that the real part of the local refractive index, which determines the mode profile, has not been altered considerably by the gain inversion, as we have previously seen in figure 3.

Transmission/reflection measurements on the probe pulse reveal that the field enhancement due to gain observed in the lower row of figure 8 leads to a complete compensation of the ohmic losses of the silver films at the emission wavelength of the dye (see figure 9). In the strongly pumped case the absorption, which for the passive structure reached a maximum value of 0.5, now becomes negative in a narrow wavelength range ($\approx$ 7 nm) around 710 nm, while the sum of the transmitted and reflected electromagnetic energy exceeds the input energy of the probe pulse. Note that, although over-compensation of losses has been achieved, the double-fishnet metamaterial does not yet act as an amplifier in transmission direction, since a significant part of the input pulse is reflected and the transmission coefficient does not exceed 1.

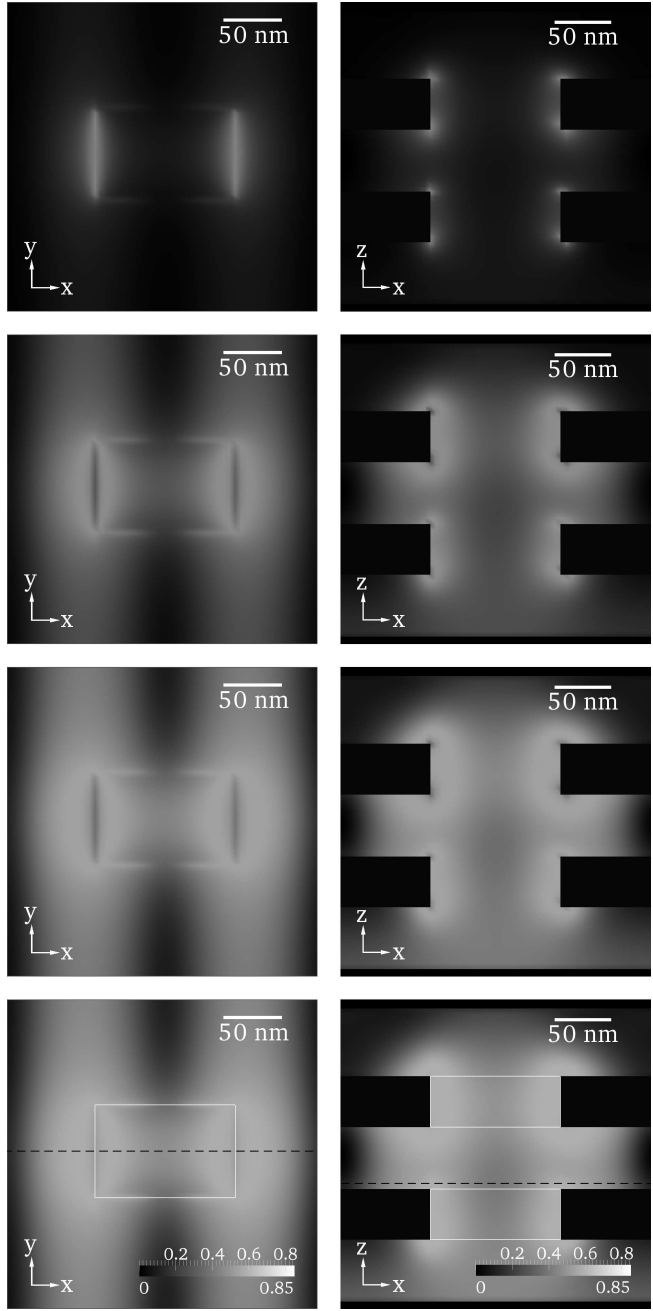

Figure 7. Snapshots of the occupation inversion $\Delta N_\mathrm{e}$ during the pump process in two perpendicular planes in a unit cell of the double fishnet. The snapshots are taken (from top to bottom) after 1.36 ps, 2.04 ps, 2.72 ps and 3.40 ps. The maximum intensity of the pump pulse is reached after approximately 2.9 ps. In the lowest row the white rectangles highlight the position of the holes and the black dashed lines show the intersection of the two planes. The rectangular black areas in the right column are intersections with regions that do not contain gain media (metal films and free space).

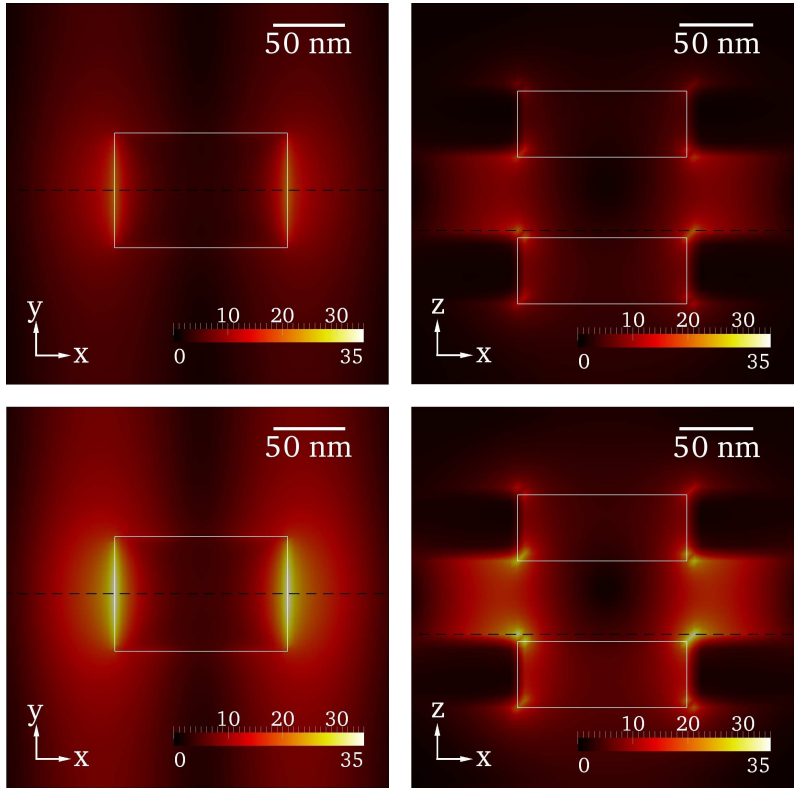

Figure 8. Electric field enhancement in two perpendicular planes of the unit cell at the probe wavelength without pumping (upper row) and with full loss compensation (lower row). This confirms that the spatial structure is retained while the amplitude is enhanced. The white rectangles highlight the position of the holes and the black dashed lines show the intersection of the two planes in each row. (Online version in colour.)

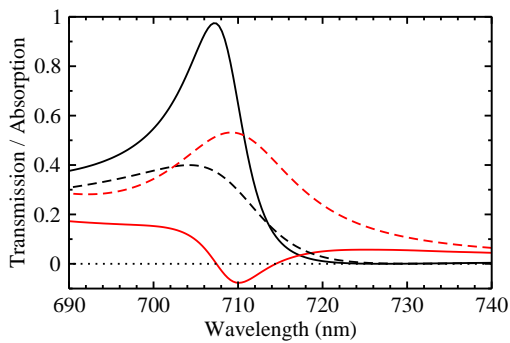

Figure 9. Transmission (black) and absorption (red) for the case without pumping (dashed lines) and for full loss compensation (solid lines). (Online version in colour.)

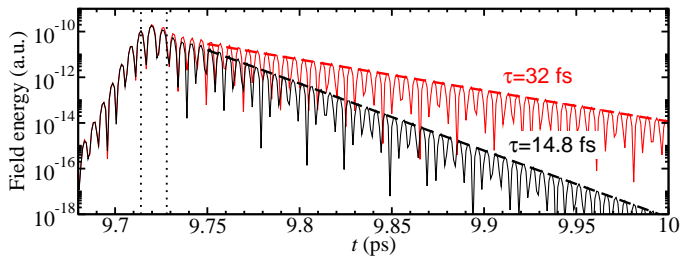

Figure 10. Temporal decay of electromagnetic energy inside the double-fishnet resonator for the case without pumping (black) and for full loss compensation (red). The vertical dotted lines indicate the FWHM of the probe pulse. (Online version in colour.)

With our time-resolved method we may also investigate the temporal decay of the electromagnetic energy inside the metamaterial. We see from figure 10 that in the passive case the electromagnetic field energy decays rapidly with time, its decay constant corresponding to an effective Q-factor of only $Q = 2\pi\tau/T \approx 40$, with $\tau \approx 14.8$ fs being the decay constant and $T \approx 2.37$ fs the period of oscillation. The decay of the field energy is due to a combination of dissipative and radiative damping. In the loss-compensation case we estimate the $Q$-factor to be enhanced by a factor of 2 ($Q \approx 85$). Notably, in the linear regime probed here, we do not observe signatures of lasing, such as gain depletion. We attribute this to the insufficient feedback (high radiation losses) of the low $Q$-factor cavity. Using a much higher density of dye molecules to enable the gain to overcome radiative as well as dissipative losses, we have been able to observe gain depletion and lasing in this same structure (results to be reported elsewhere).

### (c) Retrieved effective optical parameters

From the transmission/reflection measurements presented above we have seen that there is a wavelength region where the absorption of the double-fishnet metamaterial becomes negative. Here, we examine whether the imaginary part of the effective refractive index of the structure, retrieved with the method detailed in §3c, also becomes negative in that region. In order to avoid bianisotropy [49] the structure is deliberately placed in air instead of a thick substrate (see figure 1).

We begin by extracting the effective refractive index of the double fishnet in the absence of pumping. We see from figure 11a that there is a broad region (from around 700 nm to 740 nm) where the real part of the refractive index Re($n_{\text{eff}}$) is negative. The imaginary part Im($n_{\text{eff}}$) of the retrieved refractive index is always positive, as expected for a purely absorptive configuration. In that case the figure of merit FOM $= -n'(\lambda)/n''(\lambda)$ stays below a maximum value of around 2 (see figure 11b). The impact of a progressively increased pump intensity is also presented in figure 11. Specifically we find that for a pump intensity of 200 kV/cm the retrieved imaginary part of the effective refractive index switches sign in a wavelength region around 710 nm. This region almost exactly coincides with the region where the absorption of the metamaterial also becomes negative as we have seen in figure 9. Importantly, we see from figure 11a that here $\text{Re}(n_{\text{eff}})$ remains negative and, in fact, it even increases in magnitude. As a consequence of the zero-crossing in $\text{Im}(n_{\text{eff}})$ the figure of merit diverges at two wavelengths bounding the loss-compensation region (see dotted lines in figure 11b). For intermediate pump intensities we see from figure 11b that the FOM improves but no full loss compensation is achieved yet.

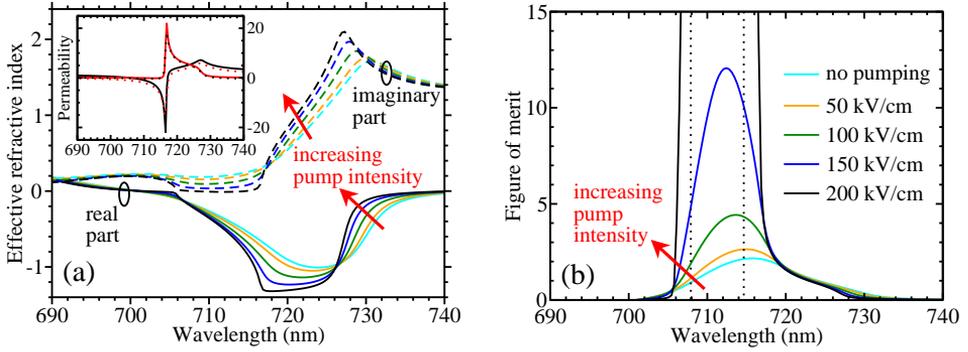

Figure 11. (a) Real and imaginary part of the retrieved effective refractive indices of the double-fishnet structure for different pump amplitudes. The peak electric field amplitude of the pump increases in steps of 50 kV/cm from no pumping (cyan line, lightest) to a maximum of 200 kV/cm (black line, darkest). The inset shows the real and imaginary parts of the effective permeability (black and red line, respectively) and the result of the Kramers-Kronig relation (black and red dotted lines) for the highest peak electric field amplitude of 200 kV/cm. (b) The figures of merit (FOMs) for the same pumping amplitudes. The vertical dotted lines indicate the exact wavelengths at which the FOM diverges. (Online version in colour.)

We have also verified that the extracted constitutive electromagnetic parameters obey causality through Kramers-Kronig relations. The inset in figure 11a presents a comparison between the values of the effective magnetic permeability as calculated by the standard retrieval method [32] and the complementary approach described in §3c. We observe that there is an excellent agreement between the values obtained by the two methods, both for the real and the imaginary part of the effective permeability. This further proves that full loss compensation is indeed achievable in the negative-index regime.

## 5. Conclusion

In summary, we have presented a self-consistent model for studying the coupled spatio-temporal plasmonic and gain dynamics in a nanostructured material with a complex three-dimensional geometry. The model incorporates the presence of free electrons in metals and bound electrons in gain materials resonantly interacting with incident optical fields. We have employed this methodology to investigate the effect of laser dye inclusions on the loss performance of an active double-fishnet metamaterial. Our model allowed us to conduct numerical pump-probe experiments and obtain an insight into the dynamical interplay between gain inversion and (coherent) plasmonic fields that led to full compensation of dissipative losses in the considered metamaterial. The presented results highlight the importance of having a good overlap of the maximum local gain coefficients with the plasmonic field enhancement in order to enable full loss compensation. Specifically for the double-fishnet metamaterial, the gain material has to be placed in the region between the two silver films. For a strongly pumped configuration, transmission/reflection measurements revealed the presence of a wavelength region where negative absorption coincided with a negative refractive index. Our work establishes a rigorous numerical model for an in-depth analysis of complex gain-enhanced plasmonic nanostructures and metamaterials.

We gratefully acknowledge financial support provided by the EPSRC, the Royal Academy of Engineering and the Leverhulme Trust.

## Appendix A. Numerical implementation of four-level system

To implement equations (2.11) and (2.12) of §2$d$ numerically, a central difference scheme for the second order polarization equations and the system of first order density equations is used. These equations are solved in each computational unit cell that contains the gain medium. Using finite differences, centred around the $n$ step with a unit time-step $\Delta t$, for the time derivatives of the polarization density, the second order polarization equations (2.12) can be expressed as

$$\frac{\mathbf{P}_i^{n+1} - 2\mathbf{P}_i^n + \mathbf{P}_i^{n-1}}{\Delta t^2} + \Gamma_i \frac{\mathbf{P}_i^{n+1} - \mathbf{P}_i^{n-1}}{\Delta t} + \omega_{0,i}^2 \mathbf{P}_i^n = -\sigma_i \Delta N_i^n \mathbf{E}^n . \qquad (A\,1)$$

Solving this equation for $\mathbf{P}_i^{n+1}$ directly gives the numerical update equation for the polarization density of the two optical transitions, $i = \text{e}$ and $i = \text{a}$.

In order to integrate the density equations (2.11) the time derivative has to be centred at $n + 1/2$, giving

$$\frac{N_3^{n+1} - N_3^n}{\Delta t} = \frac{1}{\hbar\omega_{r,\mathrm{a}}} \left( \frac{\mathbf{P}_\mathrm{a}^{n+1} - \mathbf{P}_\mathrm{a}^n}{\Delta t} + \Gamma_\mathrm{a} \frac{\mathbf{P}_\mathrm{a}^{n+1} + \mathbf{P}_\mathrm{a}^n}{2} \right) \cdot \frac{\mathbf{E}^{n+1} + \mathbf{E}^n}{2} \\ - \frac{1}{\tau_{32}} \frac{N_3^{n+1} + N_3^n}{2}, \quad \text{(A 2a)}$$

$$\frac{N_2^{n+1} - N_2^n}{\Delta t} = \frac{1}{\hbar\omega_{r,\mathrm{e}}} \left( \frac{\mathbf{P}_\mathrm{e}^{n+1} - \mathbf{P}_\mathrm{e}^n}{\Delta t} + \Gamma_\mathrm{e} \frac{\mathbf{P}_\mathrm{e}^{n+1} + \mathbf{P}_\mathrm{e}^n}{2} \right) \cdot \frac{\mathbf{E}^{n+1} + \mathbf{E}^n}{2} \\ + \frac{1}{\tau_{32}} \frac{N_3^{n+1} + N_3^n}{2} - \frac{1}{\tau_{21}} \frac{N_2^{n+1} + N_2^n}{2}, \quad \text{(A 2b)}$$

$$\frac{N_1^{n+1} - N_1^n}{\Delta t} = -\frac{1}{\hbar\omega_{r,\mathrm{e}}} \left( \frac{\mathbf{P}_\mathrm{e}^{n+1} - \mathbf{P}_\mathrm{e}^n}{\Delta t} + \Gamma_\mathrm{e} \frac{\mathbf{P}_\mathrm{e}^{n+1} + \mathbf{P}_\mathrm{e}^n}{2} \right) \cdot \frac{\mathbf{E}^{n+1} + \mathbf{E}^n}{2} \\ + \frac{1}{\tau_{21}} \frac{N_2^{n+1} + N_2^n}{2} - \frac{1}{\tau_{10}} \frac{N_1^{n+1} + N_1^n}{2}, \quad \text{(A 2c)}$$

$$\frac{N_0^{n+1} - N_0^n}{\Delta t} = -\frac{1}{\hbar\omega_{r,\mathrm{a}}} \left( \frac{\mathbf{P}_\mathrm{a}^{n+1} - \mathbf{P}_\mathrm{a}^n}{\Delta t} + \Gamma_\mathrm{a} \frac{\mathbf{P}_\mathrm{a}^{n+1} + \mathbf{P}_\mathrm{a}^n}{2} \right) \cdot \frac{\mathbf{E}^{n+1} + \mathbf{E}^n}{2} \\ + \frac{1}{\tau_{10}} \frac{N_1^{n+1} + N_1^n}{2}. \quad \text{(A 2d)}$$

These equations can be rewritten in matrix form

$$A_{kl} N_l^{n+1} = B_{kl} N_l^n + \mathbf{C}_k \cdot \left( \mathbf{E}^{n+1} + \mathbf{E}^n \right) \quad \text{(A 3)}$$

with the two matrices

$$\{A_{kl}\} = \begin{pmatrix} \Delta t^{-1} + \gamma_{32}/2 & 0 & 0 & 0 \\ -\gamma_{32}/2 & \Delta t^{-1} + \gamma_{21}/2 & 0 & 0 \\ 0 & -\gamma_{21}/2 & \Delta t^{-1} + \gamma_{10}/2 & 0 \\ 0 & 0 & -\gamma_{10}/2 & \Delta t^{-1} \end{pmatrix} \quad \text{(A 4)}$$

and

$$\{B_{kl}\} = \begin{pmatrix} \Delta t^{-1} - \gamma_{32}/2 & 0 & 0 & 0 \\ \gamma_{32}/2 & \Delta t^{-1} - \gamma_{21}/2 & 0 & 0 \\ 0 & \gamma_{21}/2 & \Delta t^{-1} - \gamma_{10}/2 & 0 \\ 0 & 0 & \gamma_{10}/2 & \Delta t^{-1} \end{pmatrix}. \quad \text{(A 5)}$$

The constant $\mathbf{C}_k$ contains a combination of new and old polarizations $\mathbf{P}^{n+1}$ and $\mathbf{P}^n$ and can be obtained by comparing equations (A 2a) to (A 2d) with the matrix equation (A 3). The decay times $\tau_{ij}$ have been replaced by decay constants $\gamma_{ij} = 1/\tau_{ij}$.

By multiplying both sides of the matrix equation (A 3) with $\{A_{kl}\}^{-1}$, update equations for the occupation densities are obtained. These update equations supplement the FDTD update equations for electromagnetic fields in form of discretized auxiliary differential equations. As $\mathbf{E}^{n+1}$ and $\mathbf{E}^n$ are not available simultaneously, the update to occupation densities $N_l^{n+1}$ in equation (A 3) has to be split into two parts: A first part, performed immediately after the $\mathbf{P}^{n+1}$ update, involves the terms containing $N_l^n$ and $\mathbf{E}^n$. The second part, requires adding terms with $\mathbf{E}^{n+1}$ and therefore has to be performed after the update of the electric field to $\mathbf{E}^{n+1}$ but before the $\mathbf{P}^{n+2}$ update.